\documentstyle [12pt]{article}

\newcommand{\cc}{\mbox{\sc C}}
\newcommand{\la}{\lambda}

\begin{document}

\begin{titlepage}
\begin{flushright}
ITP-SB-94-11 \\
MARCH 1994
\end{flushright}
\vspace{1cm}
\begin{center}
{\large \bf ALGEBRAIC ASPECTS OF BETHE-ANSATZ}
\end{center}
\vspace{1cm}
\begin{center}
{\bf L.D. Faddeev}\\
\vspace{1cm}
St.Petersburg Branch of Steklov Mathematical Institute,\\
Fontanka 27,\\
St. Petersburg 191011,\\
Russia
\\
and
\\
Research Institute for Theoretical Physics,\\
University of Helsinki,\\
Siltavuorenpenger 20C,\\
SF 00014 Helsinki,\\
Finland
\end{center}
\vspace{1cm}
\begin{abstract}
In these lectures the introduction to algebraic aspects of Bethe Ansatz is
given. The  applications to the seminal spin 1/2 XXX model
is discussed in detail and the generalization to higher spin
as well as XXZ and lattice Sine-Gordon model are indicated.
The origin of quantum groups and their appearance
in CFT models is explained. The text can be considered as a guide to the
research papers in this field.
\end{abstract}

\end{titlepage}

\section*{Preface}

During last few years I have given several short lecture courses on the
application of (algebraic) Bethe Ansatz to the integrable models of quantum
field theory in $1+1$ dimensional space-time, see \cite{1}, \cite{2}, \cite{3}.
Being essentially similar in spirit they differ in the choice of particular
topics and the technical details. This course, which was delivered at the
Institute of Theoretical Physics, University of New York at Stony Brook,
is self-contained, and do not copy \cite{1}-\cite{3}. It can be considered
as a short and non-technical introduction into the subject of integrable
models. Interested students must turn to the vast literature.
A minimal list of references is appended. A lot of references to historic
as well as research papers could be found therein.

The job of taking notes of the course and producing the lecture notes was
undertaken by Kostas Skenderis. Without his help this text would not appear.
I am very grateful to him for his devoted work.

I would like to acknowledge the hospitality which was provided to me by
Professor C.N. Yang in the Institute for Theoretical Physics, University
of New York at Stony Brook.
\newpage

\section{Lecture 1}

\subsection{Introduction.}

There are many different historical directions which led to the subject of
exactly solvable model: (1) the study of magnetic chains initiated by Bethe in
the early thirties, (2) the work of Onsanger and Baxter in 2 dimensional
classical statistical mechanics, (3) scattering theory in many body problem
with factorizable S-matrix (Berezin, Yang).  Here we will follow one more
path which again leads to exactly solvable models, namely the inverse
scattering method.

Solitons are discovered almost a hundred years ago. These are particle-like
solutions of non-linear differential equations. The prototype
equation is the KdV (Korteweg-de Vries) equation
\begin{equation}
u_t = u u_x + u_{xxx},
\end{equation}
where $u_t$ and $u_x$ denote differentiation with respect to $t$ and $x$,
respectively. It has solution which describes a localized wave which moves
without dissipation. The KdV equation resurrected in the early sixties
through the work of Kruskal and others. They are the ones who named the
particle-like solution soliton.

What is very interesting is the existence of many-soliton solutions
which opens new possibilities for the particle spectrum.
In contrast with the usual field theory where to each field corresponds
just one particle, here we have a novel situation where  one field might
generate more than one particle. This possibility might be familiar nowdays
after the emergence of the string theory but back in the early seventies
the soliton mechanism of emergence of mass spectrum was the first example
of circumvention of the perturbative paradigm.

Using the inverse scattering method, many different  theories in $1+1$
dimensions were studied and solved, see for example monograph \cite{4}.
Among them the Bose Gas model, whose equation of motion is a nonlinear
Schr\"{o}dinger (NS) equation
\begin{equation}
i \frac{\partial \Psi}{\partial t} =
- \frac{\partial^2 \Psi}{\partial x^2}
+ g |\Psi|^2 \Psi,
\end{equation}
where $\Psi(x,t)$ is a complex field and $g$ is the coupling constant,
and the Sine-Gordon model. The latter describes a relativistic real field
$\phi(x,t)$ whose equation of motion is given by
\begin{equation}
\Box \phi + \frac{m^2}{\beta} \sin \beta \phi = 0.
\end{equation}
The constants $m^2$ and $\beta$ are a mass parameter and  the
coupling constant of the model, respectively.
The solution of this model revealed a very
interesting spectrum with the mass of the particles given by the formula
\begin{equation}
m_n = \frac{16m}{\gamma} \sin \frac{n \gamma}{16},
\end{equation}
where $\gamma=8 / \beta^2$.
These studies finally led to the so-called algebraic Bethe ansatz (1978).

A natural question is what a $1+1$ dimensional  system has to do with the
real world which is 4 dimensional.
First of all, string theory teach us that 2D physics
(worldsheet) can be ultimately related with higher dimensional world
(target space).
In fact, string theories are described by conformal field theories (CFT)
which  can be viewed as ultraviolet limit of
exactly solvable models. (Alternatively, one might think the massive exactly
solvable models as deformations of CFT's).

A second example where $1+1$ physics enters comes from 4D QCD. There are
evidences that the high energy scattering is mainly 2D phenomenon (Lipatov,
Verlinde-Verlinde).

The similarities between non-linear sigma model in 2D and Yang-Mills theories
in 4D (asymptotic freedom, mass through dimensional transmutation etc.) gives
one more motivation for studying 2D models. One hopes that 2D physics can
teaches us how to tackle corresponding 4D problems.

\subsection{The XXX model.}

Consider  the one-dimensional quantum periodic chain with N sites.
We associate with each site $n$ a local spin variable
\begin{equation}
\vec{s}=\frac{1}{2} \vec{\sigma},
\end{equation}
where the components of the vector $\vec{\sigma}$ are the Pauli matrices.
The spin variables act on the  Hilbert space $h=\cc^2$.
The full Hilbert space is the tensor product of all local Hilbert spaces
\begin{equation}
{\cal H} = h^{\otimes N}.
\end{equation}
Its dimension is dim$({\cal H})=2^N$.
The spin variable act on the full Hilbert space ${\cal H}$ as follows
\begin{equation}
\vec{s}_n = I \otimes I \cdots \otimes \vec{s} \otimes \cdots I,
\end{equation}
where $I$ is the unit operator in the local Hilbert space $h$.
The periodicity of the lattice means that $\vec{s}_{n+N}=\vec{s}_n$.
The Hamiltonian of the system is given by
\begin{equation}
H= \sum_{n=1}^N \Big[ (s_n, s_{n+1}) -\frac{1}{4} \Big],
\end{equation}
where $(s_n, s_{n+1})$ is the inner product between $\vec{s}_n$ and
$\vec{s}_{n+1}$ defined in the space of two spins, namely
$\cc^2 \otimes \cc^2$.
Notice that the  Hamiltonian is manifestly negative.

Notice that we have done all possible
regularizations. The lattice spacing $\Delta$ and the finite volume $N$
provides natural ultraviolet and infrared cut-offs in
the theory whereas the choice of the quantum spin as our variable renders
the Hilbert space finite dimensional.
After we solve the theory we will wish to take the
thermodynamic limit $N \rightarrow \infty$. This is a non-trivial limit
and it is related with the theory of infinite tensor product.
However, in some cases of physical interest things are simpler.
For example, in the case of
ferromagnet the configuration consists mostly
of, say,  spin up configurations so only a finite number of spins could
be turned down.

We shall use  the permutation operator
\begin{equation}
P=\frac{1}{2} (I \otimes I + \vec{\sigma} \otimes \vec{\sigma}).
\end{equation}
If we choose as a basis in $\cc^2$ the vectors
\begin{eqnarray}
|+> = \left( \begin{array}{c}
1 \\ 0
\end{array} \right), \; \;
|-> = \left( \begin{array}{c}
0 \\ 1
\end{array} \right)
\end{eqnarray}
then the permutation operator is given by the matrix
\begin{eqnarray}
P = \left( \begin{array}{cccc}
1 & 0 & 0 & 0 \\
0 & 0 & 1 & 0 \\
0 & 1 & 0 & 0 \\
0 & 0 & 0 & 1
\end{array}
\right)
\end{eqnarray}
in the basis $|++>, |+->, |-+>, |-->.$
The Hamiltonian can be written in terms of the permutation operator as
\begin{equation}
H= \frac{1}{2} \sum_{n=1}^N ( P_{n,n+1} -1 ). \label{hamiltonian}
\end{equation}

We now introduce the Lax operators $L_{n,a}$.
Their significance will become clear
later on. They are operator which act on the local space $h_n \otimes V$
(now we use the notation $h_n$ for a local quantum space assigned to the site
$n$), where $V$ is an auxiliary space.
In our example $h=\cc^2$  is the same as $V$ but in general it
does not have to coincide with it. The $L_{n,a}$ is given by
\begin{eqnarray}
L_{n,a}(\lambda)
& = & \lambda (I_n \otimes I_a) + i (\vec{s}_n \otimes \vec{\sigma}_a)
\label{l} \\
& = & ( \lambda - \frac{i}{2} ) I + i P_{n,a} \label{lax}  \\
& = & \left( \begin{array}{cc}
\lambda + i s_n^3 & i s_n^- \\
i s_n^+ & \lambda - i s_n^3
\end{array}
\right), \label{laxop}
\end{eqnarray}
where $s^{\pm}=s^1 \pm i s^2$ and $\lambda$ is complex parameter.
The index $n$ refers to the site position
whereas the index $a$ is always an auxiliary space index. The role of the
parameter $\lambda$ will be made clear later.
In the last line we treat $L_{n,a}$ as an matrix on $V$ with coefficients
as operators in $h_n$.

The Lax operator can be interpreted as a connection on the 1D lattice.
Let $\phi_n$ be vectors from ${\cal H} \otimes V$ assigned to each lattice
site. We say that $\phi_n$ is parallel if
\begin{equation}
\phi_{n+1} = \frac{1}{\lambda} L_n \phi_n, \label{connection}
\end{equation}
(where the factor $1/ \lambda$ is a suitable rescaling), so that
the Lax operator just realizes parallel transport.
If we take the formal classical continuous limit $\Delta \rightarrow 0$,
$\hbar \rightarrow 0$ (of course after we reinstate the $\hbar$ dependence)
and define the continuous spin variable as
\begin{equation}
s(x)= \lim_{\Delta \rightarrow 0} \frac{s_n}{\Delta}; \; x=n \Delta,
\end{equation}
then using (\ref{l}) and (\ref{connection}) we get
\begin{equation}
\frac{1}{i} \phi'(x)=
\lim_{\Delta \rightarrow 0}
\frac{1}{i} \frac{\phi_{n+1} - \phi_n}{\Delta}=
\frac{S(x)}{\lambda} \phi(x) \label{inverse}
\end{equation}
where $S(x)=(\vec{s}(x),\vec{\sigma})$. But this is the equation which was
used in the inverse scattering method applied to magnet by Takhajan\cite{4}.
So (\ref{connection}) is a natural lattice and  quantum deformation of
(\ref{inverse}).

The next task is to establish commutation relations between the Lax operators.
First we introduce our notation
\begin{displaymath}
A_{ik} B_{mn} \equiv (A \otimes B)_{im|kn} \equiv (A^1 B^2 )_{im,kn},
\end{displaymath}
where $A^1=A \otimes I$ and $ B^2 = I \otimes B$.
Since $L_{n,a}$ is a $4 \times 4$ matrix there are 16 matrix elements whose
commutation relations we want to study. All these commutation relations can be
{compactly written as
\begin{equation}
R_{a_1,a_2}(\lambda - \mu) L_{n,a_1}(\lambda) L_{n,a_2}(\mu)=
L_{n,a_2}(\mu) L_{n,a_1}(\lambda)R_{a_1,a_2}(\lambda - \mu). \label{fcr}
\end{equation}
This is an equation in $V_1 \otimes V_2 \otimes h_n$. The indices $a_1$ and
$a_2$ and the variables $\lambda$ and $\mu$ are associated with
the auxiliary spaces $V_1$ and $V_2$, respectively.
The matrix $R_{a_1,a_2}$ governs the commutation relation of the matrix
elements of Lax operators. In our case it is given by
\begin{equation}
R_{a_1,a_2}(\lambda) =
\frac{1}{\lambda + i} (\lambda I_{a_1,a_2} + i P_{a_1,a_2}). \label{rmatrix}
\end{equation}
There is a nice graphical representation of (\ref{fcr}). The Lax operator can
be represented as a cross of two lines (fig. 1), one which represents the
auxiliary space index (thin line) and one which represent the site index
(thick line).
Then the graphical representation of (\ref{fcr}) is given in figure 2.
The right hand of (\ref{fcr}) is obtained from the left hand side by just
shifting the $n$-line to the other side of the cross point of the remaining
lines.

\setlength{\unitlength}{1mm}

\parbox[c]{58mm}{

\begin{picture}(58,40)

\put (20,35){\line(1,-1){20}}
\put (15,35){$a_1$}
\thicklines
\put (20,15){\line(1,1){20}}
\put (15,10){$n$}
\end{picture}
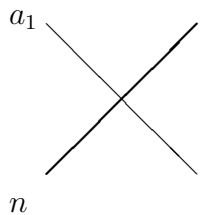
Figure 1. Graphical representation of the Lax operator $L_{n,a_1}$.\\ \\}
\ \
\parbox[c]{70mm}{

\begin{picture}(70,40)

\thinlines
\put (5,10){\line(1,1){25}}
\put (2,7){$a_2$}
\put (8,30){\line(2,-1){20}}
\put (4,30){$a_1$}
\thicklines
\put (25,34){\line(-1,-6){5}}
\put (25,35){$n$}

\put (35,20){\line(1,0){7}}
\put (35,22){\line(1,0){7}}

\thinlines
\put (45,10){\line(1,1){25}}
\put (42,7){$a_2$}
\put (48,30){\line(2,-1){20}}
\put (44,30){$a_1$}
\thicklines
\put (55,34){\line(-1,-6){5}}
\put (55,35){$n$}

\end{picture}
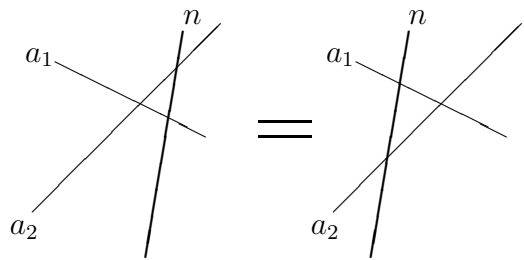
Figure 2. Graphical representation of Fundamental Commutation Relations
(\ref{fcr}).\\}

Having interpreted the Lax operator as a kind of connection along the chain,
we can introduce an operator which describes parallel transport once around
the chain. This operator is the monodromy matrix $T_a(\lambda)$
\begin{equation}
T_a(\lambda) = L_{N,a}(\lambda) L_{N-1,a}(\lambda) \cdots  L_{1,a}(\lambda).
\label{tt}
\end{equation}
The monodromy matrix satisfies commutation relations identical with the ones
satisfied by the Lax operators, namely
\begin{equation}
R_{a_1,a_2}(\lambda-\mu) T_{a_1}(\lambda) T_{a_2}(\mu)
= T_{a_2}(\mu) T_{a_1}(\lambda) R_{a_1,a_2}(\lambda-\mu). \label{monodromy}
\end{equation}
These commutation relations can be easily proven by repeatedly use of
(\ref{fcr}). The graphical representation of (\ref{monodromy}) (fig. 3) gives
an alternative (and easier) proof of (\ref{monodromy})
(the ``train'' argument).

\parbox[c]{130mm}{

\begin{picture}(130,60)

\thinlines
\put (10,40){\line(1,-1){20}}
\put (30,40){\line(1,0){25}}
\put (5,40){$a_1$}
\put (10,20){\line(1,1){20}}
\put (30,20){\line(1,0){25}}
\put (5,20){$a_2$}
\thicklines
\put (37,15){\line(0,1){30}}
\put (35,10){$N$}
\put (42,15){\line(0,1){30}}
\put (44,17){$\cdots$}
\put (50,15){\line(0,1){30}}
\put (49,10){$1$}

\put (57,30){\line(1,0){7}}
\put (57,32){\line(1,0){7}}

\thinlines
\put (70,40){\line(1,0){25}}
\put (95,40){\line(1,-1){20}}
\put (65,40){$a_1$}
\put (70,20){\line(1,0){25}}
\put (95,20){\line(1,1){20}}
\put (65,20){$a_2$}
\thicklines
\put (77,15){\line(0,1){30}}
\put (75,10){$N$}
\put (83,15){\line(0,1){30}}
\put (84,17){$\cdots$}
\put (90,15){\line(0,1){30}}
\put (89,10){$1$}

\end{picture}

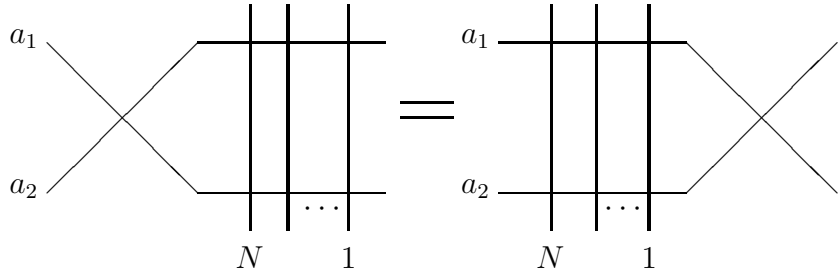
Figure 3. The ``train'' argument.
\\ }

We introduce a family of operators
\begin{equation}
F(\lambda) = {\rm Tr}_a ( T_a(\lambda) ),
\end{equation}
where the trace is over the auxiliary space. The operator $F(\lambda)$ is
acts on the full Hilbert space ${\cal H}$ (global operator).
{}From its definition
it follows that $F(\lambda)$ is a polynomial in $\la$ of degree $N$.
It follows from (\ref{monodromy}) that $F(\lambda)$ and $F(\mu)$ commute
\begin{equation}
[F(\lambda),F(\mu)]=0, \label{integrability}
\end{equation}
so there are $N$ independent quantities which mutually commute. As we
will see in a moment the Hamiltonian is among them. So we have $N$
integrals of motion in involution for a system with $N$ degrees of freedom.
In the classical case it corresponds to the
Liouville's definition of integrability.

Consider the Lax operator at the specific point $\lambda = i/2$.
{}From (\ref{lax}) we get
\begin{equation}
L_{n,a}(\lambda = i/2) = i P_{n,a}.
\end{equation}
Hence,
\begin{equation}
F(\lambda = i/2) = i^N {\rm Tr}_a (P_{N,a} P_{N-1,a} \cdots P_{1,a}).
\end{equation}
Using $P_{n,a}P_{m,a}=P_{m,n} P_{n,a}$, $P_{n,m}=P_{m,n}$ and
${\rm Tr}_aP_a=I$ we get
\begin{equation}
F(\lambda = i/2) = i^N P_{1,2} P_{2,3} \cdots P_{N,N-1}. \label{fl}
\end{equation}
But the right hand side is just the shift operator
\begin{equation}
U = P_{1,2} P_{2,3} \cdots P_{N,N-1},
\end{equation}
which has the property that,
\begin{equation}
U X_n = X_{n+1} U,
\end{equation}
for every local operator $X_n$. It follows that $U$ is just the exponential
of the momentum operator $\Pi$
\begin{equation}
U=\exp i \Pi.
\end{equation}
We now want to express the Hamiltonian in terms of the operators $F(\lambda)$.
Since
\[
\frac{d}{d \lambda} L_{n,a} = I_{n,a},
\]
we get
\begin{equation}
\frac{dF}{d \lambda} \Bigg|_{\lambda = i/2} =
i^{N-1} \sum_{j=1}^N {\rm Tr}_a
(P_{N,a} \cdots \hat{P}_{j,a} \cdots P_{1,a}),
\end{equation}
where the hat denotes omission. Comparing (\ref{fl}) and (\ref{hamiltonian})
we see that the Hamiltonian can be written as
\begin{equation}
H = \frac{i}{2} \frac{dF}{d \lambda} F^{-1}(\lambda) \Bigg|_{\lambda = i/2}
- \frac{N}{2}.
\end{equation}
Notice that since the $F(\la)$ commute for different $\la$ there is no
problem with definition of the logarithmic derivative.

Let the monodromy matrix be
\begin{equation}
T(\lambda) = \left( \begin{array}{cc}
A(\lambda) & B(\lambda) \\
C(\lambda) & D(\lambda)
\end{array}
\right).
\end{equation}
We want to diagonalize
\begin{equation}
F(\lambda) = A(\lambda) + D(\lambda).
\end{equation}
It is easy to show that
\begin{equation}
\Omega = \prod_{n=1}^N \bigotimes |+>_n.
\end{equation}
is an eigenvector of $F(\lambda)$.
Indeed, since
\begin{equation}
s^+_n |+>=0,
\end{equation}
we get that $C(\lambda) \Omega = 0$, and consequently that the matrix
$T(\lambda) \Omega$ is a triangular matrix.
Furthermore, using (\ref{laxop}), we get
\begin{eqnarray}
A(\lambda) \Omega & = & (\lambda + \frac{i}{2})^N \Omega,\\
D(\lambda) \Omega & = & (\lambda - \frac{i}{2})^N \Omega
\end{eqnarray}
so that
\begin{equation}
F(\lambda) \Omega = \big[ (\lambda + \frac{i}{2})^N
+ (\lambda - \frac{i}{2})^N \big] \Omega.
\end{equation}
Now we are going to show that the $B(\lambda)$ can be used as
spectrum raising operator. This will be done in the next lecture.
\newpage

\section{Lecture 2}

In the last lecture we have seen that the vector $\Omega$ which consist of
spin up configuration in all lattice sites is an eigenvector of $F(\la)$.
The vector $\Omega$ plays a role similar with the vacuum state of the
harmonic oscillator. To generate other states we act with $B(\la)$
on $\Omega$. In that sense the operator $B(\la)$ is a kind of raising
operator. However, the new vectors will be eigenvector of $F(\la)$ only
for specific values of $\la$'s.

Consider  the state
\begin{equation}
\Phi(\{\la\}) = B(\la_1) B(\la_2) \cdots B(\la_l) \Omega. \label{def,phi}
\end{equation}
To check if this it is an eigenvector of $A(\la)$ and $D(\la)$ we need the
commutation relation between $A(\la)$, $D(\la)$ and $B(\la)$. These are
obtained from (\ref{tt}) and are given by
\begin{eqnarray}
&\ & [ B(\la), B(\mu) ] = 0, \label{b-b}\\
&\ & A(\la) B(\mu) = \alpha(\la-\mu) B(\mu) A(\la)
+ \beta(\la-\mu)  B(\la) A(\mu), \label{a-b} \\
&\ & D(\la) B(\mu) = \gamma(\la-\mu) B(\mu) D(\la)
+ \delta(\la-\mu)  B(\la) D(\mu), \label{d-b}
\end{eqnarray}
where $\alpha(\la)= (\la -i)/ \la$ and  $\gamma(\la)=(\la +i)/ \la$. The
specific values of $\beta$ and $\delta$ are  irrelevant for our discussion.
{}From (\ref{b-b}) we see that the $\Phi$ is symmetric function of $\la$'s.
If the second term in the right hand side of (\ref{a-b}) and (\ref{d-b})
was absent then the vector $\Phi(\{\la\})$ would have been an eigenvector
of $A(\la)$ and $D(\la)$ for all values of $\la$'s. However, the presence
of these terms generates extra unwanted terms when we commute the
$A(\la)$ (or $D(\la)$) to the right. For a particular set of $\la$'s all
extra terms vanish and $\Phi(\{\la\})$ is, at the end, an eigenvector of
$F(\la)$. This set is determined by the solution of a transcendental equation.
For this particular set of $\la$'s the eigenvalue equation reads
\begin{equation}
F(\la) \Phi(\{\la\}) =
\Big[ \prod_{m=1}^l \Big( \frac{\la_m - \la + i}{\la_m - \la} \Big)
(\la +\frac{i}{2})^N
+ \prod_{m=1}^N \Big( \frac{\la_m - \la - i}{\la_m - \la}\Big)
(\la - \frac{i}{2})^N \Big] \Phi(\{\la\}). \label{eigen}
\end{equation}
Since the left hand side of (\ref{eigen}) is an analytic function of $\la$,
so has to be the right hand side. However, the right hand side has poles at
$\la = \la_l$.  All the poles are removed if the $\la$'s obey the equations
\begin{equation}
\big(\frac{\la_k + i/2}{\la_k - i/2} \big)^N =
\prod_{m=1;m \neq k}^l \frac{\la_k - \la_m + i}{\la_k - \la_m - i}
\hspace{1cm} k=1, \ldots, l. \label{cons}
\end{equation}
In fact, this is exactly the condition for the cancellation of the unwanted
extra terms  mentioned above (one can find a formal proof in \cite{5}.
Thus, self-consistency of (\ref{eigen})
automatically provides the set of $\la$'s for which it is valid.

Consider the shift operator
\begin{equation}
U = e^{i \Pi} = i^{-N} F(\la) \Big|_{\la=i/2}.
\end{equation}
We apply $U$ to $\Phi$. Since the second term in right hand side of
(\ref{eigen}) vanishes at $\la=i/2$, the spectrum of $U$ is
multiplicative and, hence, the spectrum of the momentum operator
$\Pi$ is additive
\begin{equation}
\Pi  \Phi = \sum_{m=1}^l p(\la_m) \Phi,
\end{equation}
where
\begin{equation}
p(\la_m)=\frac{1}{i} \ln \frac{\la_m + i/2}{\la_m -i/2}. \label{momentum}
\end{equation}
For $\la$ real the momentum ranges over $[0, 2 \pi]$.
The spectrum of the Hamiltonian is  also additive
\begin{equation}
H \Phi = \sum_{m=1}^l h(\la_m) \Phi,
\end{equation}
where
\begin{equation}
h(\la) = \frac{1}{2} \frac{d p(\la)}{d \la} \label{energy}
= -\frac{1}{2} \frac{1}{ \la^2 + 1/4} \le 0.
\end{equation}
The parameter $\la$ can be interpreted as the ``rapidity''.
(Recall that the energy and the momentum of a relativistic particle can be
parametrized in terms of the rapidity $\theta$ as $p=m \sinh \theta$,
$E=m \cosh \theta$.)
Eliminating the parameter $\la$ we get the dispersion relation
\begin{equation}
h(p) = \frac{1}{2} ( \cos p -1 ). \label{dispersion}
\end{equation}

The momentum of the system is quantized due to the fact that we have our
system in the finite box. This becomes manifest if we rewrite (\ref{cons})
as
\begin{equation}
e^{i p(\la_k) N} = \prod_{m=1}^l S( \la_k - \la_m), \label{quantization}
\end{equation}
where
\begin{equation}
S(\la) = \frac{\la +i}{\la -i}, \label{smatrix}
\end{equation}
is the two-particle scattering amplitude. (Compare with the free particle case
where the quantization equation is $\exp i p N = 1$.) The fact that the
$l$-particle scattering amplitude in the right hand side of
(\ref{quantization}) is expressed
in terms of two-particle one is a manifestation of the
integrability of the model.

Since the Hamiltonian is negative, the particle spectrum does not describe
physical particles. However, if we reverse the sign of the Hamiltonian
(ferromagnetic case) then we get physical particles which are called magnons.

Another observable of our system is the global spin. The spin operator is
defined as
\begin{equation}
\vec{\Sigma} = \sum_{n=1}^N \vec{s}_n.
\end{equation}
The spin operator appears in the $1/ \la$ expansion of the monodromy matrix
\begin{equation}
T(\la) = \la^N \Big[I + \frac{\vec{\Sigma} \cdot \vec{\sigma}_a}{\la} +
{\cal O}(\frac{1}{\la^2}) \Big].
\end{equation}
The ${\cal O}(1 / \la^2)$ term is related with Yangian symmetries\cite{6}.
Consider now the commutation relations (\ref{tt}) in the limit
$\la \rightarrow \infty$, $\mu$ fixed. It follows that
\begin{equation}
[\frac{1}{2} \vec{\sigma}_a + \vec{\Sigma}, T_a (\mu)] = 0,
\end{equation}
which means that the monodromy matrix is invariant under combined
rotations in  the full quantum and auxiliary space.
One have in particular the following relations
\begin{equation}
[\vec{\Sigma}, F(\la)] = 0, \label{rotation}
\end{equation}
which implies rotational invariance of the Hamiltonian $H$,
\begin{equation}
[\Sigma_3, B(\la)] = - B(\la), \label{s3b}
\end{equation}
which means that the operator $B(\la)$ turns down one spin and
\begin{equation}
[\Sigma_+, B(\la)] = A(\la) - D(\la). \label{s+b}
\end{equation}
We evidently have
\begin{equation}
\Sigma_3 \Omega = \frac{N}{2} \Omega,
\end{equation}
so that (\ref{s3b}) leads to
\begin{equation}
\Sigma_3 \Phi = (\frac{N}{2}-l) \Phi, \label{s3}
\end{equation}
where $l$ is the number of $B$-operators acting on $\Omega$
(see (\ref{def,phi})). There are many comments in order here.
First of all notice the difference between the odd and even case. When $N$ is
even we get integer spin states, whereas when it is odd we get
semi-integer spin states.
In particular the ground state can only appear in the  even case.
So contrary to the usual intuition, the nature of spectrum is different for
the $N$ even and $N$ odd case even in the limit $N \rightarrow \infty$.
Next we can prove using (\ref{s+b}) and the Bethe ansatz equations
(\ref{cons}) that all $\Phi$ states are highest weight states, i.e.
\begin{equation}
\Sigma_+ \phi = 0. \label{highest}
\end{equation}
To get the rest of the states we act with $\Sigma_-$. Since $\Sigma_-$
commutes with the Hamiltonian $H$ all the descendants have the same energy.
Furthermore, equation (\ref{s3}) and (\ref{highest}) imply that $l \leq N/2$.

One can easily check that (\ref{eigen}) is invariant under complex conjugation.
This means that for every complex solution $\la$ of (\ref{eigen}),
$\bar{\la}$ is also a solution.
In addition all $\la$'s are different (Pauli principle)\cite{7}.
This can be inferred
by examining again the self-consistency of (\ref{cons}). If two $\la$'s
were equal then (\ref{cons}) would have a double pole. So we would need
$(l+1)$ equations in order to remove all poles which is an over-determined
system (we have only $l$ unknowns). In fact one can prove that in
the large $N$ limit the $(l+1)$ equations are incompatible. So the $\la$'s
ought to be different.

We want to study the vacuum state. From (\ref{s3}) we infer that $l=N/2$.
In addition, more detailed investigation shows that all $\la$'s are real.
We take the logarithm in both sides in (\ref{cons}) and we use a branch of
the logarithm in  the form
\begin{equation}
\frac{1}{i} \ln \Big( \frac{\la + ia}{\la - ia} \Big)
= \pi - \arctan \frac{\la}{a}.
\end{equation}
Equation (\ref{cons}) becomes
\begin{equation}
N \arctan 2 \la_j = \sum_{l=1}^{N/2} \arctan (\la_j - \la_l) + \pi Q_j,
\label{logar}
\end{equation}
where $Q_j$ are quantum numbers which parametrize the $\la$'s.
They are integers or half-integers. One can prove
that $\la$ is a monotonic function of $Q_j$ so  that
$Q_j \rightarrow Q_j^{\rm max}$ in the limit
$\la \rightarrow \infty$.
In fact, for finite $N$ the range of $Q$ is
\begin{equation}
-\frac{N}{4} + \frac{1}{2} \leq Q \leq  \frac{N}{4} - \frac{1}{2},
\end{equation}
so there $N/2$ $Q$'s.
This means that $Q$ takes all the values from $-Q_{\rm max}$ to
$Q_{\rm max}$, so actually we have $Q_j=j$.
The ground state consists of a completely filled Dirac sea.

We now wish to take the thermodynamic limit $N \rightarrow \infty$.
We introduce a new variable $x=j/N$. In the thermodynamic limit
$\la_j \rightarrow \la(x)$. Assuming regularity of the function
$\la(x)$ one can replace the sum in (\ref{logar}) by an integral within an
${\cal O} (1/N^2)$ error:
\begin{equation}
\arctan 2 \la(x) = \pi x + \int_{-1/4}^{1/4} dy \arctan
\big[\la(x)- \la(y)\big]. \label{cont,eq}
\end{equation}
This a non-linear equation but it can be reduced to a linear one as we will
now show.
Differentiating equation (\ref{cont,eq}) w.r.t. x we get
\begin{equation}
\frac{2 \la'(x)}{1+4\la^2} = \pi + \la'(x)
\int \frac{1}{ 1+ (\la(x) -\la(y))^2} dy. \label{step}
\end{equation}
We introduce the density of states, $\rho(\la)$, defined as
\begin{equation}
\rho(\la)=\frac{1}{\la'(x)}. \label{rho}
\end{equation}
This is a natural definition since $dx = \rho(\la) d \la$.
Using (\ref{rho}), equation (\ref{step}) becomes a linear integral equation
for the density of states $\rho(\la)$:
\begin{equation}
\frac{2}{1+4\la^2} = \pi \rho(\la) +
\int_{- \infty}^{\infty} \frac{1}{ 1+ (\la -\mu)^2} \rho(\mu) d\mu.
\label{integr}
\end{equation}
This equation was obtained by L. Hulthen can be solved by means of
Fourier transform.

The ground state energy in the thermodynamic limit is given by
\begin{eqnarray}
E &=& \sum h(\la_l) \rightarrow N \int h(\la(x)) dx \nonumber \\
&=& N \int_{-\infty}^{\infty} h(\la) \rho(\la) d\la.
\end{eqnarray}
It can be shown that it is equal to
\begin{equation}
E_0 = - N \ln 2 = - \ln 2^N.
\end{equation}
This value is related with the residual entropy at $T=0$.

Let us now consider excitations. In the $N$-even case the first excitation
 appears when $l=N/2-1$ (triplet state) and it consists of two vacancies(holes)
in the Fermi sea. Each of them carries spin 1/2.
The parameters $Q$ are modified
\begin{equation}
Q_j = j + \theta(j-j_1) + \theta(j-j_2),
\end{equation}
where $\theta$ is the Heaviside function. The parameters $Q_j$ have two
jumps at the position of the holes.
Equation (\ref{cont,eq}) is modified accordingly
\begin{equation}
\arctan 2 \la(x) = \pi x
+ \frac{\pi}{N} \big[ \theta(j-j_1) + \theta(j-j_2) \big] +
\int_{-1/4}^{1/4} dy \arctan
\big[\la(x)- \la(y)\big]. \label{cont,eq1}
\end{equation}
and, hence,
\begin{equation}
\frac{2}{1+4\la^2} = \pi \rho(\la)
+ \frac{\pi}{N} \big[\delta(\la-\la_1) + \delta(\la-\la_2)\big]
+ \int_{- \infty}^{\infty} \frac{1}{ 1+ (\la -\mu)^2} \rho(\mu) d\mu,
 \label{integr1}
\end{equation}
where $\la_i, \hspace{0.3cm} i=1, 2$, are the rapidities  for the holes.
Thus, the density of states is also modified
\begin{equation}
\rho(\la;\la_1,\la_2) = \rho_{\rm vac}(\la)
+ \frac{1}{N} \big( \sigma(\la-\la_1) + \sigma(\la-\la_2) \big)
\end{equation}
where $\rho_{\rm vac}(\la)$ is the vacuum density of states and $\sigma(\la)$
is determined by the equation
\begin{equation}
\pi \sigma(\la) +
\int_{- \infty}^{\infty} \frac{1}{ 1+ (\la -\mu)^2} \sigma(\mu) d\mu
= - \pi \delta(\la).
\end{equation}
The energy of the system is the sum of the vacuum energy plus the
contribution from the holes
\begin{equation}
E = E_{\rm vac} + {\cal E}(\l_1) + {\cal E}(\l_2),
\end{equation}
where
\begin{equation}
{\cal E}(\la) = \int_{-\infty}^{\infty} h(\mu) \sigma(\mu-\la) d\mu.
\end{equation}
Similarly, the relative momentum $k(\la)$ is given by
\begin{equation}
k(\la) = \int_{-\infty}^{\infty} p(\mu) \sigma(\mu -\la) d\mu.
\end{equation}
It ranges from 0 to $\pi$. The relative energy ${\cal E}(\la)$ and momentum
$k(\la)$ can be interpreted as the energy and momentum of the
physical excitation.
The dispersion relation is very simple
\begin{equation}
{\cal E}(k) = \frac{\pi}{2} \sin k.
\end{equation}
The 1-particle excitation will come from the $N$-odd sector.

The scattering matrix $S_{\rm ex}$ for the triplet excitation is determined
by the equation
\begin{equation}
\ln S_{\rm ex}(\la) = \int \ln S(\mu) \sigma(\mu-\la) d \mu,
\end{equation}
where $S(\mu)$ is given in (\ref{smatrix}). It is given by
\begin{equation}
S_{\rm ex}(\la) = \frac{f(\la)}{f(-\la)},
\end{equation}
where
\begin{equation}
f(\la) = \frac{\Gamma(1/2 + i\la/2)}{\Gamma(1 + i\la/2)} =
\frac{\Gamma(s)}{\Gamma(1/2 + s)}; \hspace{0.5cm}
s=\frac{1}{2} + i\frac{\la}{2}.
\end{equation}
It is interesting to note that the function $|f(\la)|^2$ appears also in the
representations of $SL(2,R)$ as a Harish-Chandra factor.

The singlet solution can be described in a similar fashion, but one make
use of the complex solution of BA equations (see \cite{5}).

Up to now we have only considered the antiferromagnetic case
(Hamiltonian negative definite).
Let us briefly consider now the ferromagnetic case
(Hamiltonian positive definite).
The physical vacuum coincide with the Fock space vacuum state
(compare with the antiferromagnetic case where the physical vacuum is
constructed by filling the Dirac sea).
In the infinite volume limit $N \rightarrow \infty$ equation
(\ref{cons}) does not yield quantization of momentum.
The energy (\ref{dispersion}) becomes positive after the change of sign of
the Hamiltonian and corresponds to physical particles.
The operators $\Sigma_+$ and $\Sigma_-$ cease to exist and only
\begin{equation}
Q=\Sigma_3 - \frac{N}{2},
\end{equation}
makes sense.
Thus the $SU(2)$ symmetry breaks down. The spectrum has also bound states
with the dispersion law
\begin{equation}
{\cal E}_M(p) = \frac{1}{2M+1} (1-\cos p),
\end{equation}
where $M$ is an eigenvalue of charge $Q$.

Most of the techniques described so far apply to other integrable models
as well. In general to define a model we need a group $G$,
a representation $\rho$, and a spectral parameter.
In our case the spectral parameter was defined on the complex plane.
However, it may also be defined on a strip (non-degenerate rational case),
or on a torus (elliptic case). We must introduce deformation parameters
$\gamma$ and $\kappa$ to define these two cases.
Thus one can classify all the integrable models in terms of the quartet
\[
(G, \rho, \gamma, \kappa)
\]
For example the NS model corresponds to general spin, whereas the Sine-Gordon
to general spin plus one deformation.
\newpage

\section{Lecture 3}

In this lecture we will consider higher spin generalizations of the spin-1/2
XXX model. Instead of spin-1/2 variables at each lattice site we now have
spin-$S$ variables.
The local Hilbert space is modified accordingly
\begin{equation}
h_n = \cc^{2S+1}.
\end{equation}
One can define a Lax operator $L_n(\la)$ in the same way as earlier,
\begin{equation}
L_n(\la) = \left( \begin{array}{cc}
\lambda + i s_n^3 & i s_n^- \\
i s_n^+ & \lambda - i s_n^3
\end{array} \label{laxpair}
\right),
\end{equation}
but now acts on $\cc^{2S+1} \otimes \cc^2$.
The commutation relations between the $L_{n,a}$'s are given by
\begin{equation}
R_{a_1,a_2}(\lambda - \mu) L_{n,a_1}(\lambda) L_{n,a_2}(\mu)=
L_{n,a_2}(\mu) L_{n,a_1}(\lambda)R_{a_1,a_2}(\lambda - \mu), \label{fcr1}
\end{equation}
where $R_{a_1,a_2}$ is given again by (\ref{rmatrix}).

\setlength{\unitlength}{1mm}

\parbox[c]{130mm}{

\begin{picture}(130,40)

\thinlines
\put (35,10){\line(1,1){25}}
\put (30,5){$1/2$}
\put (38,30){\line(2,-1){20}}
\put (32,32){$1/2$}
\thicklines
\put (45,34){\line(-1,-6){5}}
\put (45,35){$S$}

\put (65,20){\line(1,0){7}}
\put (65,22){\line(1,0){7}}

\thinlines
\put (75,10){\line(1,1){25}}
\put (70,5){$1/2$}
\put (78,30){\line(2,-1){20}}
\put (72,32){$1/2$}
\thicklines
\put (95,34){\line(-1,-6){5}}
\put (95,35){$S$}

\end{picture}
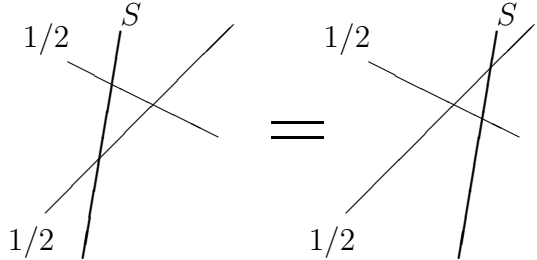
Figure 4. Graphical representation of (\ref{fcr1}).
The thin lines represent ``spin-1/2'' lines, whereas the
thick ones are ``spin-S'' lines.\\}

The monodromy matrix $T(\la)$ is defined similarly
with the spin-1/2 case.
The vector $\Omega = \prod \bigotimes \omega_n$ is again an eigenvector
of the operator $F(\la)={\rm Tr}_a T(\la)$, where $\omega_n$ is a
highest weight spin state
\begin{equation}
s_n^+ \omega_n = 0; \hspace{1cm} s_n^3 \omega_n = S \omega_n.
\end{equation}
Using again the same ansatz for the general eigenvector of $F(\la)$
\begin{equation}
\Phi(\{\la\}) = B(\la_1) B(\la_2) \cdots B(\la_l) \Omega, \label{ansatz}
\end{equation}
we get the Bethe equations
\begin{equation}
\big(\frac{\la_k + i S}{\la_k - i S} \big)^N =
\prod_{m=1;m \neq k}^l \frac{\la_k - \la_m + i}{\la_k - \la_m - i}
\hspace{1cm} k=1, \ldots, l. \label{trans}
\end{equation}
If $\{\la\}$ is a solution of (\ref{trans}) then  $\Phi(\{\la\})$ is an
eigenvector of $F(\la)$.
However, it is not clear at all how to get a local Hamiltonian.
Since the auxiliary space is different from the local Hilbert space we can not
use the same tricks we used in the spin-1/2 case where we had
$L_{n,a} = P_{n,a}$ at some specific value of $\la$.
In order to circumvent the problem we change the auxiliary space
to $V=\cc^{2S+1}$.
So, from a $S \otimes 1/2$ representation we go to a $S \otimes S$
representation.
The new Lax operator $L_{n,f}$ is defined on $\cc^{2S+1} \otimes \cc^{2S+1}$
and is called the fundamental Lax operator\cite{8}. This Lax operator is
graphically represented by a cross of two S-lines (thick lines).
In order to find its explicit form we have to solve simultaneously
the equations which are graphically represented by the figures 5 and 6.

\parbox[c]{130mm}{

\begin{picture}(130,40)

\thicklines
\put (35,10){\line(1,1){25}}
\put (32,7){$S$}
\put (38,30){\line(2,-1){20}}
\put (34,30){$S$}
\thinlines
\put (45,34){\line(-1,-6){5}}
\put (43,37){$1/2$}

\put (65,20){\line(1,0){7}}
\put (65,22){\line(1,0){7}}

\thicklines
\put (75,10){\line(1,1){25}}
\put (72,7){$S$}
\put (78,30){\line(2,-1){20}}
\put (74,30){$S$}
\thinlines
\put (95,34){\line(-1,-6){5}}
\put (93,37){$1/2$}

\end{picture}
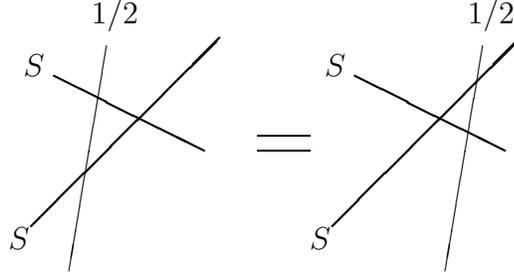
Figure 5. The S-S-1/2 equation.\\}

\parbox[c]{130mm}{

\begin{picture}(130,40)

\thicklines
\put (35,10){\line(1,1){25}}
\put (32,7){$S$}
\put (38,30){\line(2,-1){20}}
\put (34,30){$S$}
\put (45,34){\line(-1,-6){5}}
\put (45,35){$S$}

\put (65,20){\line(1,0){7}}
\put (65,22){\line(1,0){7}}

\put (75,10){\line(1,1){25}}
\put (72,7){$S$}
\put (78,30){\line(2,-1){20}}
\put (74,30){$S$}
\put (95,34){\line(-1,-6){5}}
\put (95,35){$S$}

\end{picture}
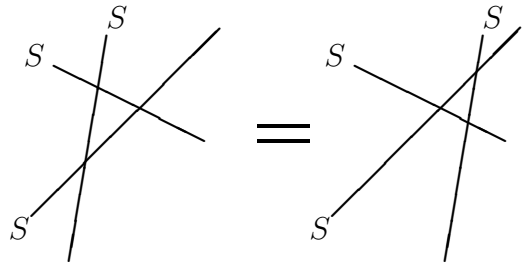
Figure 6. The S-S-S equation.\\}

Solution to these equation exists and it is unique.
The fundamental monodromy matrix is defined as
\begin{equation}
T_f = L_{n,f} \cdots L_{n,f},
\end{equation}
and is represented as in figure 7.

\parbox[c]{130mm}{

\begin{picture}(130,30)

\thicklines
\put (40,15){\line(1,0){30}}
\put (45,10){\line(0,1){10}}
\put (50,10){\line(0,1){10}}
\put (55,10){\line(0,1){10}}
\put (60,10){\line(0,1){10}}
\put (65,10){\line(0,1){10}}

\end{picture}
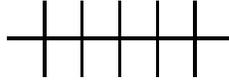
Figure 7. Graphical representation of the fundamental monodromy matrix $T_f$.
\\}

We wish to prove that the trace of the fundamental monodromy matrix commutes
with the trace of the auxiliary monodromy matrix
\begin{equation}
[{\rm Tr}_f T_f, {\rm Tr}_a T_a] = 0. \label{ttcom}
\end{equation}
The first one will yield the local conservation laws whereas the second
is the one which the Bethe ansatz gives.
Having a solution of the equations described in figures 5 and 6, the ``train
argument'' (fig. 8) immediately shows that (\ref{ttcom}) is satisfied.

\parbox[c]{130mm}{

\begin{picture}(130,60)

\thinlines
\put (10,40){\line(1,-1){20}}
\put (30,20){\line(1,0){25}}
\thicklines
\put (30,40){\line(1,0){25}}
\put (10,20){\line(1,1){20}}

\thicklines
\put (37,15){\line(0,1){30}}
\put (42,15){\line(0,1){30}}
\put (44,17){$\cdots$}
\put (50,15){\line(0,1){30}}
\put (57,30){\line(1,0){7}}
\put (57,32){\line(1,0){7}}

\thinlines
\put (70,40){\line(1,0){25}}
\put (95,40){\line(1,-1){20}}

\thicklines
\put (95,20){\line(1,1){20}}
\put (70,20){\line(1,0){25}}
\put (77,15){\line(0,1){30}}
\put (83,15){\line(0,1){30}}
\put (84,17){$\cdots$}
\put (90,15){\line(0,1){30}}

\end{picture}

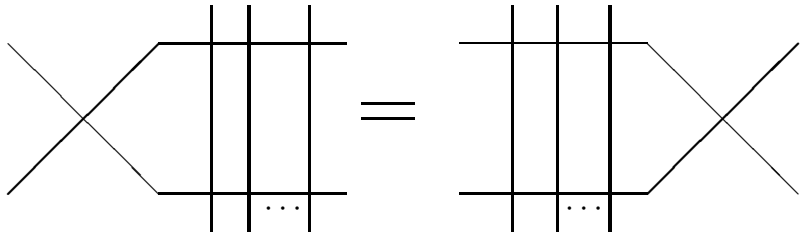
Figure 8. The ``train argument'' for equation (\ref{ttcom}).
\\ }

Let us describe the algebraic structure underlying the Bethe ansatz.
We propose the following mnemonic rule.
Let us suppose that we have an associative algebra ${\cal A}$,
and an operator $R$ on ${\cal A} \otimes {\cal A}$
which satisfies a universal Yang-Baxter equation
\begin{equation}
R_{12} R_{13} R_{23} = R_{23} R_{13} R_{12}, \label{yb}
\end{equation}
defined on  ${\cal A} \otimes {\cal A} \otimes {\cal A}$. The indices indicate
to which of ${\cal A}$'s the operator $R$ belongs.
For example $R_{13}$ means that
the operator $R$ belongs to  the first and third  ${\cal A}$.
Let $\rho(s,\la)$ be a representation of the algebra ${\cal A}$ characterized
by a discrete index $s$ (the spin in our case) and a complex variable
$\la$ (the spectral variable). A representative example is  the loop algebra
where a representation is given by $X_{a,n}=T^a \la^n$, where $T^a$ are the
spin matrices (however it is not our case).
We define
\begin{equation}
(\rho(s_1,\la_1) \otimes \rho(s_2,\la_2))R = R^{s_1 s_2}(\la_1,\la_2).
\end{equation}
We assume that $R^{s_1 s_2}(\la_1,\la_2)$ depends only on the difference
$(\la_1 -\la_2)$.
Consider the case $h_n = \cc^{2S+1}, V=\cc^2$, and apply (\ref{yb}) to
\begin{equation}
\rho(a_1,\la) \rho(a_2,\mu) \rho(n,\sigma),
\end{equation}
where the $a$-indices refer to the auxiliary space and the $n$ to the
local Hilbert space,as usual. Equation (\ref{yb}) becomes
\begin{equation}
R_{a_1,a_2}(\la-\mu) R_{a_1,n}(\la-\sigma) R_{a_2,n}(\mu-\sigma) =
R_{a_2,n}(\mu-\sigma) R_{a_1,n}(\la-\sigma) R_{a_1,a_2}(\la-\mu). \label{rrl}
\end{equation}
If we identify $R_{a,n}(\la)$ with $L_{n,a}(\la)$ then (\ref{rrl}) is just
equation (\ref{fcr1}). The value $\la=0$ is assumed to be the specific
point where $L_{n,a}$ coincides with the permutation operator $P_{n,a}$.
We permute the indices in (\ref{yb}) according to the permutation
\[
\left(
\begin{array}{ccc}
1 & 2 & 3 \\
3 & 1 & 2
\end{array}
\right).
\]
This yields
\begin{equation}
R_{12} R_{32} R_{31} = R_{31} R_{32} R_{12}. \label{ybp}
\end{equation}
Acting in (\ref{ybp}) with
\begin{equation}
\rho(n_1,\la) \rho(n_2,\mu) \rho(a,\sigma),
\end{equation}
we get
\begin{equation}
R_{n_1,n_2}(\la-\mu) R_{a,n_2}(\sigma-\mu) R_{a,n_1}(\sigma-\la) =
R_{a,n_1}(\sigma-\la) R_{a,n_2}(\sigma-\mu) R_{n_1,n_2}(\la-\mu), \label{rll}
\end{equation}
or with the identification of $R_{a,n}(\la)$ with $L_{n,a}(\la)$,
\begin{equation}
L_{n_1,a} (\sigma-\la) L_{n_2,a} (\sigma-\mu) R_{n_1,n_2}(\la-\mu) =
R_{n_1,n_2}(\la-\mu) L_{n_2,a} (\sigma-\mu) L_{n_1,a} (\sigma-\la).
\end{equation}
This is the exact form of the equation given graphically in figure 5.

It is now straightforward to find the Hamiltonian of a spin-$S$ system.
It is given by
\begin{equation}
H  = \sum_n H_{n,n+1} = \sum_n f_s(\vec{s}_n \cdot  \vec{s}_{n+1}),
\end{equation}
where
\begin{equation}
H_{n_1,n_2}
= \frac{1}{i} \frac{d}{d\la} R_{n_1,n_2}(\la)\Big|_{\la=0} P_{n_1,n_2},
\end{equation}
and $f_s(\vec{s}_n \cdot  \vec{s}_{n+1})$ is polynomial of degree $2s$.
For example for a spin-1 system we have
\begin{equation}
f_1(x) = x - x^2.
\end{equation}

We now generalize our results to the case of the arbitrary complex spin
variable $S$ (Verma moduli). The spin matrices can be expressed in terms of
Bose creation  and annihilation operators $\psi^*$ and $\psi$ which satisfy
\begin{equation}
[\psi^*, \psi] = -1, \label{bose}
\end{equation}
as (Holstein-Primakov)
\begin{eqnarray}
s^+_n &=& \psi_n^* (2 S - \psi_n^* \psi_n)^{1/2}, \label{s+} \\
s^-_n &=& (2 S - \psi_n^* \psi_n)^{1/2} \psi_n, \label{s-} \\
s^3_n &=& \psi_n^* \psi_n - S.
\end{eqnarray}
If $\psi_n^* \psi_n$ is an integer then we are dealing with a finite subspace
since necessarily
\begin{equation}
\psi_n^* \psi_n \leq 2S.
\end{equation}
When $S$ is complex $s^+, s^-$ and $s^3$ form a representation of $SL(2,C)$
and not of $SU(2)$.
In the limit $S \rightarrow \infty$ the spin matrices limit to
\begin{eqnarray}
\lim_{s \rightarrow \infty} s^+_n &=& \sqrt{2S} \psi_n^*, \\
\lim_{s \rightarrow \infty} s^-_n &=& \sqrt{2S} \psi_n, \\
\lim_{s \rightarrow \infty} s^3_n &=& -S.
\end{eqnarray}
Let us find the $S \rightarrow \infty$ limit of the Lax pair given in
(\ref{laxpair}),
\begin{eqnarray}
i\frac{L_n(\la)}{S} \sigma_3 &=& \frac{i}{S}
\left( \begin{array}{cc}
\la+is^3_n & is^-_n \\
is^+_n & \la-is^3_n
\end{array}
\right) \sigma_3 \\
&=& I +
\left( \begin{array}{cc}
i\la /S & \psi_n / \sqrt{2S} \\
-\psi^*_n/ \sqrt{2S} & -i\la /S
\end{array}
\right). \label{inftylimit}
\end{eqnarray}
Consider now the continuum limit $\Delta n \rightarrow x$.
Let $\Delta = 1/S$. The creation and annihilation operators are of order
$\sqrt{\Delta}$. To see this note that the discretized version of a delta
function is
\begin{equation}
\delta(x-y) \rightarrow \frac{\delta_{mn}}{\Delta},
\end{equation}
so that  (\ref{bose}) becomes
\begin{equation}
[\frac{\psi_n^*}{\sqrt{\Delta}}, \frac{\psi_m}{\sqrt{\Delta}}] =
- \frac{\delta_{mn}}{\Delta},
\end{equation}
which proves our assertion.
Defining the continuum $\psi$-field as
\begin{equation}
\psi(x) = \frac{\psi_m}{\sqrt{\Delta}},
\end{equation}
equation (\ref{inftylimit}) becomes
\begin{equation}
\frac{i L(x, \la)}{S} = I + \Delta
\left( \begin{array}{cc}
\la & \psi(x) \\
\psi^*(x) & -\la
\end{array}
\right),
\end{equation}
where we have absorbed a factor of $i$ in the parameter $\la$. But this exactly
the Lax operator for nonlinear Sch{\"o}dinger (NS) equation\cite{4}.
Hence the NS model belongs to same class of model with the XXX magnetic
chains. It is recovered in the $S \rightarrow \infty$ limit of the latter.
The momentum and the energy in the NS model are given by $k$ and $k^2$,
respectively. Let us see if we can get these values from the XXX model.
We have seen that the momentum in ferromagnetic case is given by
\begin{equation}
p(\la) = \frac{1}{i} \ln \Big( \frac{\la + iS}{\la - iS} \Big).
\end{equation}
In the limit $S \rightarrow \infty$, $p(\la)$ goes as $\la/S$.
Furthermore, the energy limits to
\begin{equation}
E = \frac{2S}{S^2 + \la^2} \sim  \frac{2}{S} - \frac{2 \la^2}{S^3}.
\label{enns}
\end{equation}
Identifying $k$ with an appropriately  rescaled $\la$ (and subtracting
the constant from (\ref{enns})) we see that the $S \rightarrow \infty$
limit of XXX model correctly reproduces the energy and momentum spectrum.
Let us stress that here we differ from others, who prefer to view the NS
model as a limit of the spin-1/2 XXZ model.

The spin-1/2 XXZ model is defined by the Hamiltonian
\begin{equation}
H = \sum_n \big( s_n^x s_{n+1}^x + s_n^y s_{n+1}^y + J s_n^z s_{n+1}^z \big),
\end{equation}
where $J$ is some constant different from 1. We will treat the XXZ model
as a $q$-analogue of the XXX model.

The notion of $q$-deformation was introduced by Gauss in the last century.
He defined the $q$-deformation of $x$ as
\begin{equation}
x_q = \frac{q^x - q^{-x}}{q - q^{-1}}.
\end{equation}
The limit $q \rightarrow 1$ gives back the original value
\begin{equation}
\lim_{q \rightarrow 1} x_q = x.
\end{equation}
If we introduce a new variable $\gamma$,
\begin{equation}
q = e^{i \gamma},
\end{equation}
then $x_q$ is given in terms of the new variable $\gamma$ by
\begin{equation}
x_q = \frac{\sinh \gamma x}{\sin \gamma}.
\end{equation}
One can now deform the Lax operator of the XXX model as
\begin{equation}
L(\la,\gamma) =
\left( \begin{array}{cc}
\sinh [\gamma ( \la + i s^3)] & is^- \sin \gamma \\
is^+ \sin \gamma & \sinh [\gamma ( \la - i s^3)]
\end{array}
\right).
\end{equation}
The deformed $R$ matrix looks as follows
\begin{equation}
R = \left( \begin{array}{cccc}
\sinh [\gamma(\la +i)] & 0 & 0 & 0 \\
0 & \sinh \gamma \la & i\sin \gamma & 0 \\
0 & i\sin \gamma & \sinh \gamma \la & 0 \\
0 & 0 & 0 & \sinh [\gamma(\la +i)]
\end{array}
\right).
\end{equation}
Note that we have changed the normalization w.r.t. (\ref{rmatrix}).
If we impose the fundamental commutation relations (\ref{fcr}),
we discover that the spin matrices obey deformed algebraic relations\cite{9},
\begin{equation}
[s^3, s^{\pm}] = \pm s^{\pm}
\end{equation}
\begin{equation}
[s^{+}, s^-] =  \frac{\sin (2 \gamma s^3)}{\sin \gamma}.
\end{equation}
This algebraic relations define what we now call quantum group.
Actually, a better name would be deformed Lie algebra.
In the mathematical literature the notion of deformation of algebraic
structure is well defined.
One starts by making the  structure constants $C_{ij\cdots}$ $t$-dependent,
where $t$ is a deformation parameter. The algebraic structure determined
by the deformed structure constants is, in general, nonequivalent to  the
original one since some of the properties of the corresponding algebra
are changed after deformation. A well known example
is quantum mechanics. It has been proven that quantum mechanics is
stable deformation of classical mechanics (the deformation variable is
the $\hbar$). In this sense, one can use the word ``quantum''
instead of  the word ``deformation''.  Drin'field was the first who used the
term ``quantum'' in this sense and thus coined the term ``quantum group''
\cite{10}.
\newpage

\section{Lecture 4}

In the last lecture we have seen that the XXZ$_S$ model can be viewed as a
$q$-deformation of the XXX$_S$ model. We have constructed the Lax operator
$L(\la)$ of the XXZ model as a $q$-deformation of Lax operator of the XXX
model and we have seen that it obeys the fundamental commutation relations,
\begin{equation}
R(\la-\mu) L^1(\la) L^2(\mu) = L^2(\mu) L^1(\la) R(\la-\mu), \label{core}
\end{equation}
only when the spin matrices obey certain relations. These considerations led
to the discovery of the quantum groups.

Now we shall elaborate on this in more detail.
It will be convenient to perform a similarity transformation in all
entries in (\ref{core})
\begin{equation}
L(\la) \rightarrow Q(\la) L(\la) Q^{-1}(\la)
\end{equation}
and
\begin{equation}
R(\la-\mu) \rightarrow
Q^1(\la) Q^2(\mu) R(\la-\mu) Q^1(\la)^{-1} Q^2(\mu)^{-1}, \label{rsim}
\end{equation}
where
\begin{equation}
Q(\la) =
\left( \begin{array}{cc}
q^{-i\la/2} & 0 \\
0 & q^{i\la/2}
\end{array}
\right).
\end{equation}
In matrix form $Q^1(\la) Q^2(\mu)$ is given by
\begin{equation}
Q^1(\la) Q^2(\mu) = \left( \begin{array}{cccc}
q^{-i(\la+\mu)/2} & 0 & 0 & 0 \\
0 & q^{-i(\la-\mu)/2} & 0 & 0 \\
0 & 0 & q^{i(\la-\mu)/2} & 0 \\
0 & 0 & 0 & q^{i(\la+\mu)/2} \\
\end{array}
\right),
\end{equation}
and only the middle block of it acts non-trivially in (\ref{rsim}).
So the new $R$-matrix
depends on the difference $(\la-\mu)$. Explicitly we have now
\begin{equation}
L(\la,\gamma) =
\left( \begin{array}{cc}
\sinh [\gamma ( \la + i s^3)] & is^- \sin \gamma e^{\la\gamma} \\
is^+ \sin \gamma e^{-\la\gamma} & \sinh [\gamma ( \la - i s^3)]
\end{array}
\right), \label{rep}
\end{equation}
and,
\begin{equation}
R = \left( \begin{array}{cccc}
\sinh [\gamma(\la +i)] & 0 & 0 & 0 \\
0 & \sinh \gamma \la & i\sin \gamma e^{\la\gamma} & 0 \\
0 & i\sin \gamma e^{-\la\gamma} & \sinh \gamma \la & 0 \\
0 & 0 & 0 & \sinh [\gamma(\la +i)]
\end{array}
\right).
\end{equation}
We introduce a new variable
\begin{equation}
x = \exp \gamma \la.
\end{equation}
In terms of this variable the Lax operator decomposes as follows
\begin{equation}
L = x L_+ - \frac{1}{x} L_-, \label{ldec}
\end{equation}
where
\begin{equation}
L_+ = \left( \begin{array}{cc}
q^{H/2} & (q - q^{-1})s^- \\
0 & q^{-H/2}
\end{array}
\right), \label{10}
\end{equation}
\begin{equation}
L_- = \left( \begin{array}{cc}
q^{H/2} & 0 \\
- (q - q^{-1})s^+ & q^{-H/2}
\end{array}
\right), \label{explicit-l-}
\end{equation}
and we introduce a notation $q^{H/2}=\exp i \gamma s^3$.
Similarly, the $R$-matrix can be written as
\begin{equation}
R(\la) = xR_+ - \frac{1}{x} R_-, \label{rdec}
\end{equation}
where
\begin{equation}
R_+ = \left( \begin{array}{cccc}
q & 0 & 0 & 0 \\
0 & 1 & (q-q^{-1}) & 0 \\
0 & 0 & 1 & 0 \\
0 & 0 & 0 & q
\end{array}
\right), \label{explicit-r}
\end{equation}
and,
\begin{equation}
R_- = P R_+^{-1} P. \label{property}
\end{equation}
$P$ is the permutation operator.
We now substitute (\ref{ldec}) and (\ref{rdec}) into (\ref{core}).
In each side of the resulting equation  seven different powers of $x$ and $y$
($y= \exp \gamma \mu$) appear, namely
\[
\{x^2, y^2, x^2y^2, x^2/y^2, y^2/x^2, 1/(x^2y^2), 1\}.
\]
This implies  seven different equations which relate $R_\pm$ with $L_\pm$.
So, for example, equating the coefficients of ``1'' we get
\begin{equation}
R_+ L_-^1 L^2_+ + R_- L_+^1 L^2_- = L^2_+ L_-^1 R_+ + L^2_- L_+^1 R_-.
\label{1eq}
\end{equation}
The other six equations are given by
\begin{eqnarray}
R_{\pm (\mp)} L^1_\pm L^2_\pm &=& L^2_\pm L^1_\pm R_{\pm (\mp)} \label{4eq} \\
R_+ L^1_+ L^2_- &=& L^2_- L^1_+ R_+ \label{rlrl} \\
R_- L^1_- L^2_+ &=& L^2_+ L^1_- R_-, \label{last}
\end{eqnarray}
where the notation $R_{\pm (\mp)}$ means that we have to consider all
four cases, namely plus or minus in $R$, plus sign in both $L$'s,
and plus or minus in $R$, minus sign in both $L$'s.
So, equation (\ref{4eq}) is, in fact, four equations.
However, only three of the seven equations are independent.
We choose as our independent equations the following three
\begin{eqnarray}
R L^1_+ L^2_+ &=& L^2_+ L^1_+ R, \label{1ind} \\
R L^1_- L^2_- &=& L^2_- L^1_- R, \label{2ind}  \\
R_+ L^1_+ L^2_- &=& L^2_- L^1_+ R_+,\label{3ind}
\end{eqnarray}
where $R$ can be either $R_+$ or $R_-$.
Let us illustrate  how we can derive the rest of the equations
starting from the above three. In particular we derive the following equation
\begin{equation}
R_- L^1_+ L^2_+ = L^2_+ L^1_+ R_-, \label{apote}
\end{equation}
starting from the equation
\begin{equation}
R_+ L^1_+ L^2_+ = L^2_+ L^1_+ R_+. \label{para}
\end{equation}
We multiply both sides with the permutation operator $P$.
The left hand side gives
\begin{eqnarray}
P R_+ L^1_+ L^2_+ &=& R^{-1}_- P L^1_+ L^2_+ \nonumber \\
&=& R^{-1}_- L^2_+ L^1_+ P, \label{lhs}
\end{eqnarray}
where in the first step we used (\ref{property}).
Similarly, the right hand side yields
\begin{equation}
P L^2_+ L^1_+ R_+ = L^1_+ L^2_+ R^{-1}_- P. \label{rhs}
\end{equation}
Equations (\ref{lhs}) and (\ref{rhs}) imply (\ref{apote}).
All but equation (\ref{1eq}) can be derived similarly.
Equation (\ref{1eq}) can be checked using the so-called Hecke property
\begin{equation}
R_+ - R_- = (q - \frac{1}{q})P,
\end{equation}
which can be checked explicitly from (\ref{explicit-r}) and (\ref{property}).

The relations (\ref{4eq})-(\ref{last}) can be taken as the defining
relations for the $q$-deformation of the Lie algebra $SL(2)$ expressed in
terms of its generators $q^H, s_+$ and $s_-$, entering in (\ref{10}),
(\ref{explicit-l-})\cite{11}.
They are homogeneously quadratic. The matrices $R_\pm$
play the role of the corresponding structure constants.
The Yang-Baxter (Y-B) relation
\begin{equation}
R_+^{12} R_+^{13} R_+^{23} = R_+^{23} R_+^{13} R_+^{12}, \label{yang-baxter}
\end{equation}
and analogous relations involving $R_-$ guarantee that higher order relations
follow from the quadratic ones. (One can derive (\ref{yang-baxter}) from
(\ref{rrl}) for $n=a_3$ in an exactly the same way as we find
(\ref{4eq})-(\ref{last})
from (\ref{core})). The relation (\ref{yang-baxter}) guarantees that
(\ref{1ind})-(\ref{3ind}) are the only relations defining the deformed
algebra $SL(2)_q$.
Indeed, following the two paths in the diagram
\begin{equation}
L^1 L^2 L^3 =
\begin{array}{ccc}
\ & L^1 L^3 L^2 \rightarrow L^3 L^1 L^2 & \ \\
\nearrow & \ & \searrow \\
\searrow & \ & \nearrow \\
\ & L^2 L^1 L^3 \rightarrow L^2 L^3 L^1 & \
\end{array}
= L^3 L^2 L^3.
\end{equation}
we get the cubic relation
\begin{equation}
R^{123} L^3 L^2 L^1 (R^{123})^{-1} =
R^{321} L^3 L^2 L^1 (R^{321})^{-1}, \label{cubic-y-b}
\end{equation}
where $R^{123}$ and $R^{321}$ are the left hand side and the right hand side
of the relation (\ref{yang-baxter}), respectively. So, due to this relation
the relation (\ref{cubic-y-b}) is empty. Very general theory-categorical
considerations show that if the cubic relations are absent (so if the
Y-B relation (\ref{yang-baxter}) holds) then no new relations will occur
in any order of $L$.

Consider now the operator
\begin{equation}
\hat{R}=P R_+.
\end{equation}
The Yang-Baxter relation takes the form
\begin{equation}
\hat{R}_{12} \hat{R}_{23} \hat{R}_{12} =
\hat{R}_{23} \hat{R}_{12} \hat{R}_{23}, \label{braid}
\end{equation}
and the Hecke relation is
\begin{equation}
\hat{R}^2 = I + (q-\frac{1}{q}) \hat{R}. \label{hecke}
\end{equation}
This is a $q$-deformation for the transposition generators of the symmetric
group $\sigma^{ik}$, realized through the permutation operator $P$, i.e.,
\begin{equation}
P_{12} P_{23} P_{12} = P_{23} P_{12} P_{23}, \label{perm}
\end{equation}
and
\begin{equation}
P^2_{ik} = I. \label{id}
\end{equation}
Indeed, we have
\begin{equation}
\hat{R}\Big|_{q=1} = P,
\end{equation}
so that (\ref{perm}) and (\ref{id}) are the $q \rightarrow 1$ limit
of (\ref{braid}) and (\ref{hecke}), respectively.

The Y-B relations (\ref{yang-baxter}) are characteristic of the Braid group
of Artin. The representation (\ref{rep}) which we associate with the
Lie algebra $SL(2)_q$ can be generalized to $SL(N)_q$.
The $R$-matrix will be then a triangular
matrix $N^2 \times N^2$ and the Hecke relation still holds.
Other classical Lie algebras lead to their own $R$-matrix and some
generalization of the Hecke relation.

The $SL(N)$ $q$-deformed Lie algebra is defined in terms of the corresponding
$R$-matrix by the relations (\ref{4eq})-(\ref{last}), where matrices of
generators $L_\pm$ are triangular and the diagonal elements are arranged in
such a way that
\begin{equation}
l^+_{ii}  l^-_{(N-i),(N-i)} = 1. \label{div}
\end{equation}
This gives the definition of the $q$-deformed Lie algebra $SL(N)_q$
(or $A_{N-1}$) corresponding to $SL(N)$. One can give a similar definition
of the $q$-deformed Lie algebras corresponding to all classical
series $B, C$ and $D$, see \cite{11}.

After this general discussion we return to the $SL(2)_q$ algebra.
We redefine $R_+$ by multiplying it with $q^{-1/2}$,
\begin{equation}
R_+ \rightarrow q^{-1/2} R_+ =
\left( \begin{array}{cccc}
q^{1/2} & 0 & 0 & 0 \\
0 & q^{-1/2} & (q-q^{-1})q^{-1/2} & 0 \\
0 & 0 & q^{-1/2} & 0 \\
0 & 0 & 0 &  q^{1/2}
\end{array}
\right),
\end{equation}
and consider its blocks as the spin 1/2 representation of the matrix $L_+$.
We see that the $q$-deformation of the Pauli matrices are given by
\begin{equation}
q^{H/2} = \left( \begin{array}{cc}
q^{1/2} & 0 \\
0 & q^{-1/2}
\end{array}
\right),
\end{equation}
\begin{equation}
s^- = \left( \begin{array}{cc}
0 & 0 \\
q^{-1/2} & 0
\end{array}
\right),
\end{equation}
and,
\begin{equation}
s^+ = \left( \begin{array}{cc}
0 & q^{-1/2} \\
0 & 0
\end{array}
\right).
\end{equation}
Let us note that here the structure constants ($R$-matrix) appear in the
fundamental representation rather than in the adjoint one.

Important property of quantum group is a comultiplication operation,
which corresponds to the addition of spins in the $q=1$ limit.
Let $L_\pm'$ and $L_\pm''$ be two independent (commuting) set of
generators. Then their matrix products $L_\pm'L_\pm''$ satisfy the
same relation as each of the $L$'s.. Indeed, we have
\begin{equation}
R (L_+'L_+'')^1 (L_+'L_+'')^2 = (L_+'L_+'')^2 (L_+'L_+'')^1 R.
\end{equation}

One can combine the generators of quantum group into one matrix instead of two,
introducing a matrix $L$,
\begin{equation}
L=L_+L_-^{-1}. \label{gauss}
\end{equation}
It satisfies the equation
\begin{equation}
L^1 R_-^{-1} L^2 R_- = R_+^{-1} L^2 R_+ L^1. \label{clas}
\end{equation}
This equation can be proven by using manipulations similar to the ones
we have used earlier. Let us see, for example, how the left hand side arises.
We start from equation (\ref{rlrl}) and multiply from the left with
$L^1 R_-^{-1}$
and from the right with $R_-^{-1} (L_-^1)^{-1} (L_-^2)^{-1} R_-$ .
Then the right hand side of (\ref{rlrl}) yields the left hand side of
(\ref{clas}).
The relation (\ref{gauss}) is a kind of Gauss decomposition for
the matrix $L$, since the diagonal of $L$ is divided by (\ref{div}).

The classical limit $q \rightarrow 1$ or equivalently
$\gamma \rightarrow 0$ is more transparent in terms of $L$.
The $R_+$ matrix becomes for small $\gamma$
\begin{equation}
R_+ = I + i \gamma r_+ + {\cal O}(\gamma^2),
\end{equation}
where
\begin{equation}
r_+ = \left( \begin{array}{cccc}
\frac{1}{2} & 0 & 0 & 0 \\
0 & -\frac{1}{2} & 2 & 0 \\
0 & 0 & -\frac{1}{2} & 0 \\
0 & 0 & 0 & \frac{1}{2}
\end{array}
\right)
\end{equation}
Similarly, one can define a matrix $r_-$ from the expansion of $R_-$.
Then the difference between $r_+$ and $r_-$ is a Casimir operator,
\begin{equation}
C \equiv r_+ - r_- = \vec{\sigma}\otimes\vec{\sigma} =
\left( \begin{array}{cccc}
1 & 0 & 0 & 0 \\
0 & -1& 2 & 0 \\
0 & 2 & -1& 0 \\
0 & 0 & 0 & 1
\end{array}
\right) \label{c-matrix}
\end{equation}
We expand also the matrix $L$,
\begin{equation}
L = I + \gamma l + {\cal O}(\gamma^2).
\end{equation}
Let us keep the Planck constant $\hbar$, so that $q=e^{i\gamma\hbar}$.
The fundamental commutation relations imply that
\begin{equation}
l^1 l^2 - l^2 l^1 = \hbar [C, l^2].
\end{equation}

It is evident, that we got the relations of Lie algebra $SL(2)$ written
in terms of the ``structure constant'' $C$.
The usual form
\begin{equation}
[l^a, l^b] = i \hbar \epsilon^{abc} l^c,
\end{equation}
is obtained if one introduces $l^a$ as
\begin{equation}
l = \sum_a l^a \sigma_a.
\end{equation}
In terms of the universal enveloping algebra, generated by $l^a$ or $L_\pm$,
we can say that it was the comultiplication which was deformed and the
multiplication, which was left intact.

The opposite deformation takes place for the dual object, $q$ deformation
Lie group, corresponding to a given Lie algebra. Let us deformed it also
in terms of matrix elements - coordinates on the group manifold -
$T=\|t_{ij}\|$. The following relation
\begin{equation}
R T^1 T^2 = T^2 T^1 R, \label{qgroup}
\end{equation}
make coordinates non-commutative, introducing a new non-commutative
multiplication.
The comultiplication
\begin{equation}
T', T'' \rightarrow T = T' T'',
\end{equation}
is the same as in the classical case $q=1$.

It is possible to combine the generators $L$ and $T$ to define a bigger
algebra. For that the following relation between $T$ and $L$ suffices
\begin{equation}
T^2 L^1 = R_+ L^1 R_-^{-1} T^2. \label{joint}
\end{equation}
which corresponds to the left action of Lie algebra on its Lie group.
The system of equations (\ref{clas}), (\ref{qgroup}) and (\ref{joint})
defines the cotangent bundle $(T^*G)_q$.
In the semi-classical limit we have
\begin{equation}
[T^2, l^1] = C T^2.
\end{equation}

The set of relations (\ref{clas}), (\ref{qgroup}) and (\ref{joint}) is
covariant with respect to the left shifts
\begin{equation}
T \rightarrow T S, \hspace{0.5cm}
L \rightarrow S L S^{-1},
\end{equation}
but then $S$ is necessarily quantized, namely it obeys
\begin{equation}
R S^1 S^2 = S^2 S^1 R.
\end{equation}
Note the matrix elements of $S$ commute with the ones of $L$ and $T$.
\newpage

\section{Lecture 5}

Let us describe in more detail the system of relations
(\ref{clas}), (\ref{qgroup}) and (\ref{joint})
which defines a ``quantum top''. As we will see later,
it is the quantum top which drives the zero mode of the WZNW model.
The phase space of the classical top consists of coordinates $g$
which are group elements and momenta $\omega$ which are elements of
the corresponding Lie algebra.
The Lagrangian of the system is given by
\begin{equation}
L = {\rm Tr}\dot{g} g^{-1} \omega - \frac{1}{2} {\rm Tr}\omega^2,
\label{lag}
\end{equation}
and the basic Poisson brackets are
\begin{eqnarray}
\{g^1, g^2\} &=& 0, \label{poisson1} \\
\{\omega^1, \omega^2\} &=& [C, \omega^2], \label{poisson2} \\
\{\omega^1, g^2\} &=& C g^2, \label{poisson3}
\end{eqnarray}
where $C=\vec{\sigma}\otimes\vec{\sigma}$ and, as before
\begin{eqnarray}
g^1 &=& g \otimes I \\
g^2 &=& I \otimes g \hspace{0.5cm} {\rm etc.}
\end{eqnarray}
Using the equations of motion
\begin{eqnarray}
\dot{\omega} &=& 0 \\
\dot{g} &=& \omega g
\end{eqnarray}
we get for the time development of the system
\begin{equation}
g(t) = \exp [\omega(0) t] g(0). \label{time}
\end{equation}

In the quantum case our system of commutation relations from
the Lecture 4 is
\begin{eqnarray}
L^1 R_-^{-1} L^2 R_- &=& R_+^{-1} L^2 R_+ L^1, \label{qpoisson1} \\
R T^1 L^2 &=& T^2 T^1 R, \label{qpoisson2} \\
T^2 L^1 &=& R_+ L^1 R_-^{-1} T^2. \label{qpoisson3}
\end{eqnarray}
In the limit $\gamma \rightarrow 0$ we identify $g$ with $T$ and $\omega$
with $l$, where $l$ is the first term in the $\gamma$ expansion of $L$,
\begin{equation}
L = I + \gamma l + \cdots.
\end{equation}
Then the quantum commutation relations (\ref{qpoisson1}), (\ref{qpoisson2})
and (\ref{qpoisson3}) limit to (\ref{poisson1}), (\ref{poisson2})
and (\ref{poisson3}), respectively. ($C = r_+ - r_-$, see eq.(\ref{c-matrix})).
The time evolution of the system is described by the equation
\begin{equation}
T(n) = L^n T(0),
\end{equation}
where the discrete index $n$ plays the role of time.
The interpretation of $T(n)$ as a time-evolved coordinate is supported by
the fact that the pair $(T(n), L)$ obeys the same quantum commutation
relations as the pair $(T(0), L)$.

Let us now consider the Wess-Zumino-Novikov-Witten (WZNW) model.
The degree's of freedom are group elements $g(x,t)$ and the left
and right currents $j_\mu^L = \partial_\mu g g^{-1}$ and
$j_\mu^R = \partial_\mu g^{-1} g$, respectively.
The model is defined on a cylinder $R^1 \otimes S^1$ or the sphere $S^2$
depending whether we are in Minkowski or Euclidean picture.
The left and right currents  separately generate a Kac-Moody algebra.
These are two independent algebras since left currents commute right currents.
{}From now on we restrict our attention to the left sector.
Let
\begin{equation}
l = j_0 + j_1.
\end{equation}
Then the fact that $l$ generates a Kac-Moody algebra is expressed
through the following Poisson bracket between the $l$'s
\begin{equation}
\{l^1(x), l^2(y)\} = \gamma [l^2(y), C] \delta(x-y) + \gamma C \delta'(x-y),
\label{kacmoody}
\end{equation}
where $\gamma$ is the coupling constant and is related to the level
$k_{\rm cl}$ of the Kac-Moody algebra by
\begin{equation}
k_{\rm cl} = \frac{\pi}{\gamma}.
\end{equation}
Here we write $k_{\rm cl}$ because quantum mechanically the level might
renormalize.

The quantum lattice picture for this looks as follows.
Consider a chain with an operator $L_n$ attached to each site.
Let $R_{\pm(n-m)}$ be
\begin{equation}
R_{\pm(n-m)} = \left\{ \begin{array}{ll}
I & \mbox{if $n \neq m$} \\
R_\pm & \mbox{if $n=m$}
\end{array}
\right.
\end{equation}
We impose the following set of commutation relations for $L_n$
\begin{equation}
L_n^1 R_{-(n-m-1)}^{-1} L_m^2 R_{-(n-m)} =
R_{+(n-m)}^{-1} L_m^2 R_{+(n-m+1)} L_n^1. \label{deltader}
\end{equation}
It is easy to see that the shift of arguments in two $R$'s
generates a derivative of a delta function in the classical limit.
Indeed for $\hbar \rightarrow 0$, $\Delta \rightarrow 0$
\begin{equation}
R_{\pm(m-n)} \rightarrow I + \gamma \delta_{mn} r_\pm,
\end{equation}
and,
\begin{equation}
L_n  \rightarrow I + \Delta l(x).
\end{equation}
(Recall that the Lax operator is a kind of connection on the 1D lattice, so
\begin{equation}
L_n \sim \stackrel{{\textstyle \leftarrow}}{\exp} \int_\Delta l(x) dx,
\end{equation}
where $\stackrel{{\textstyle \leftarrow}}{\exp} \int$ means a path ordered
integral.) \newline
Furthermore,
\begin{equation}
\frac{1}{\Delta} \delta_{mn} \rightarrow \delta(x-y),
\end{equation}
and,
\begin{equation}
\frac{1}{\Delta^2} (\delta_{n,m+1} - \delta_{nm}) \rightarrow \delta'(x-y).
\end{equation}
Combining the terms of order $\Delta^2$ in (\ref{deltader}) we get
(\ref{kacmoody}) with the $\delta'$ term present.
Quantum corrections make the following connection of deformation
parameter $q$ and Kac-Moody level $k$:
\begin{equation}
q = \exp \frac{i \pi}{k+2}.
\end{equation}

Equation (\ref{deltader}) is, in fact, three equations
\begin{eqnarray}
L^1_n L^2_n R_- &=& R_+^{-1} L^2_n L^1_n \hspace{0.5cm} (m=n),
\label{algebra1} \\
L^1_n  L^2_{n+1} &=& L_{n+1}^2 R_+ L_{n}^1 \hspace{0.5cm} (m=n+1),
\label{algebra2}
\end{eqnarray}
\begin{equation}
[L_{n}^1, L_{m}^2] = 0 \hspace{0.5cm} {\rm for} \; |m-n| \geq 2.
\label{algebra3}
\end{equation}

It is amazing that the our system still ``remembers'' the Virasoro
algebra.
Indeed, let ${\cal A}_N$ be the quantum structure generated by the Lax
operators $L_n$ in a lattice with $N$ lattice sites. Then one can
consider a new quantum structure ${\cal A}_{N'}$ generated by products
of some neighboring $L$'s, where $N'<N$.
Then ${\cal A}_{N'} \hookrightarrow {\cal A}_N$.
In other words if, for example,
$L_1, L_2, L_3, L_4, L_5$ obey the commutation relations (\ref{deltader})
so does the set $L'_1=L_1L_2, L'_2=L_3$ and $L'_3=L_4L_5$.
Hence, the density of points in the lattice is irrelevant. This is a kind of
reparametrization invariance in the lattice.

Consider now the local field $g(x)$ for the WZNW model.
In terms of the holonomy
\begin{equation}
u(x) = \stackrel{{\textstyle \leftarrow}}{\exp}
\int_0^x l(x) dx,
\end{equation}
and the corresponding holonomy for the right current $v(x)$ it can be
expressed as
\begin{equation}
g(x) = v(x) g(0) u(x),
\end{equation}
where $g(0)$ is a zero mode of the local field.
We want to find the lattice analogue of this expression.
We introduce two vertex operator $u_n$ and $v_n$,
\begin{equation}
u_n = L_n L_{n-1} \cdots L_1,
\end{equation}
and an analogue expression for $v_n$ made from right sector $L$'s.
Then,
\begin{equation}
g_n = v_n g(0) u_n.
\end{equation}
The vertex operator $u_N$ is actually equal to the monodromy matrix
$M_N=L_N L_{N-1} \cdots L_1$.
It satisfies the fundamental commutation relations
\begin{equation}
M^1 R_-^{-1} M^2 R_- = R_+^{-1} M^2 R_+ M^1. \label{m-c-r}
\end{equation}
which coincides with that of the $q$-deformed algebra (\ref{qpoisson1}).
The derivation follows from the algebra (\ref{algebra1})-(\ref{algebra3}).
The commutation relations (\ref{m-c-r}) does not depend on the lattice spacing
$\Delta$ and presumably stay intact in the proper continuum limit.
Thus the quantum Lie group naturally enters the conformal field
theory as a monodromy of the local current.

It can be shown now, that the local field $g_n$ is commutative
\begin{equation}
g_n^1 g_n^2 = g_n^2 g_n^1; \hspace{0.5cm} n \neq m,
\end{equation}
and periodic
\begin{equation}
g_{n+N} = g_n,
\end{equation}
if the pair $(g(0), M)$ constitutes the top, i.e. the following commutation
relations hold
\begin{equation}
R g^1(0) g^2(0) = g^2(0) g^1(0) R,
\end{equation}
and
\begin{equation}
g^2(0) M^1 = R_+ M^1 R_-^{-1} g^2(0).
\end{equation}
This means, that the $q$-deformed top dynamical system is a complete
system of zero modes of WZNW model and the spectrum of this latter
field-theoretical models patterned by the spectrum of the former
finite-dimensional.
Indeed, the Hilbert space of the WZNW model can be written as
\begin{equation}
{\cal H}_{\rm WZNW} = \sum_j {\cal H}_j \otimes {\cal H}_j,
\end{equation}
which is a sum of tensor squares of the irreducible representation ${\cal H}_j$
of the Kac-Moody algebra of spin $j$, $j=0, 1/2, \ldots k/2$.
An analogue formula holds for the Hilbert space of the top, namely
\begin{equation}
{\cal H}_{\rm top} = \sum_j V_j \otimes V_j,
\end{equation}
where $V_j$ are finite dimensional representations of the $q$-deformed Lie
algebra $SL(2)_q$. Moreover the embedding of $SL(2)_q$ into Kac-Moody
described above allows to state, that
\begin{equation}
{\cal H}_j = V_j \otimes {\cal H}_0,
\end{equation}
thus separating the contribution of the zero modes and oscillator degrees
of freedom. This formula gives natural definition of fusion rules
\begin{equation}
{\cal H}_i \hat{\otimes} {\cal H}_j = V_i \otimes V_j \otimes {\cal H}_0.
\end{equation}
Note that the factor ${\cal H}_0$ appears only once. These fusion rules
are based on a new comultiplication law between the Kac-Moody and the
quantum algebra\cite{12}.

Let us illustrate  the relations among the various
algebras generated by a single Lie algebra $g$ with a diagram:
\[
\begin{array}{cccccccccl}
\ & \ & \ & \ & g & \ & \ &\  & \ & {\rm Level}\ 0  \\
\ & \ & \ & \swarrow &\ & \searrow & \ & \ & \ \\
CS & \sim & KM_k & \ &\ & \ & g_q & \ & \ & {\rm Level}\ 1\\
\ & \swarrow & \ & \searrow & \ & \swarrow & \ & \searrow & \  \\
? & \ & \ & \ & KM_{k,q} & \ & \ & \ & g_{q,k} & {\rm Level}\ 2
\end{array}
\]
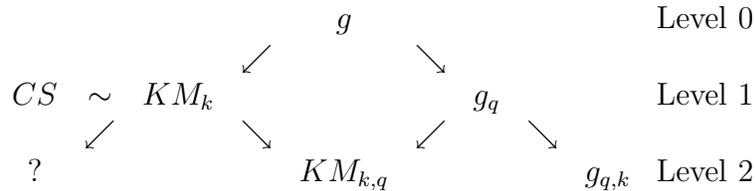
Figure 9. Relations between the algebras generated by the Lie algebra $g$.
\newpage
\noindent
At the level 1 we have either an affine infinite dimensional algebra
algebra $KM_k$ or a finite dimensional but deformed algebra $g_q$.
Both of them are parametrized by one parameter, namely the level $k$
for the Kac-Moody algebra and the deformation parameter $q$ for the
quantum algebra. As we have seen these two algebras are intimately
related. Furthermore, the Kac-Moody algebra is related with a
Chern-Simon theory (CS).
At the level 2 we can have a local Lie algebra in 2 dimensions,
an infinite deformed Lie algebra, or a further deformed finite dimensional
quantum algebra. For the last two options we have natural candidates,
namely the deformed Kac-Moody algebra $KM_{k,q}$ and the elliptic
Sklyanin algebra $g_{q,k}$, respectively.  However, the relation
between them has not been completely understood yet\cite{13}.
For the first option
there exist no natural candidate yet. Maybe a double-loop algebra is
a possible candidate but this not clear at the moment. We stop our
speculations here.

Let us now briefly show how massive models fit into our framework.
We consider the Sine-Gordon model (SG) for concreteness.
The Lax operator is the one of the XXZ model:
\begin{equation}
L(\la,\gamma) = \frac{1}{\sin \gamma}
\left( \begin{array}{cc}
\sinh [\gamma ( \la + i s^3)] & is^- \sin \gamma \\
is^+ \sin \gamma & \sinh [\gamma ( \la - i s^3)]
\end{array}
\right).
\end{equation}
However, the realization of $s^\pm, s^3$ is different.
We parametrize them in terms of real canonical fields $\pi_n$ and $\phi_n$,
\begin{equation}
[\pi_n, \phi_n] = -i I.
\end{equation}
The explicit form of the realization is the following
\begin{eqnarray}
s^- &=& \frac{1}{2 \kappa \sin \gamma} e^{-i \pi/2}
(1+ \kappa^2 e^{-2i\phi})e^{-i \pi/2}, \\
s^+ &=& (s^-)^\dagger,\\
s^3 &=& \phi.
\end{eqnarray}
It follows that the fundamental commutation relations are satisfied.
In the limit $\Delta \rightarrow 0$ and with $\kappa = m \Delta$
 one recovers the Lax operator for the Sine-Gordon model with mass $m$.
Therefore, the SG model really belongs to
the same class with the XXZ models which in turn is a deformation
of the XXX model.
Here we again differ from others who have studied the SG starting from
the XYZ model of spin 1/2.

However, if we try to repeat the same analysis as we did for the XXX or
the XXZ model we run into a problem.
The lower left matrix element of $L_n$ has no zero eigenvalue.
To get around this problem we consider the product of two  Lax
operators
\begin{equation}
L_n L_{n-1} = \left( \begin{array}{cc}
A & B \\
C & D
\end{array}
\right).
\end{equation}
Indeed, $L_n L_{n-1}$  becomes an upper diagonal matrix when applied to
a suitable chosen vacuum state $\Omega$. Furthermore, the state $\Omega$
is an eigenvector of $A$ and $D$,
\begin{eqnarray}
A \Omega &=& a(\la) \Omega, \\
D \Omega &=& d(\la) \Omega
\end{eqnarray}
The Bethe equation reads
\begin{eqnarray}
\Big( \frac{a(\la)}{d(\la)}\Big)^{N/2} &=&
\Big( \frac{ \cosh (\la + \omega - i\gamma/2)
\cosh (\la - \omega - i\gamma/2)}{
\cosh (\la + \omega + i\gamma/2) \cosh (\la - \omega - i\gamma/2)}
\Big)^{N/2} \label{bethesg} \\
&=& \prod_\mu \frac{\sinh (\la -\mu +i\gamma)}{\sinh (\la -\mu -i\gamma)},
\end{eqnarray}
where $\omega$ satisfies the equation
\begin{equation}
2 \cosh 2\omega = m^2 + \frac{1}{m^2}.
\end{equation}
Let us compare this result with the Bethe equation for the XXZ model.
There the left hand side has the form
\[
\frac{\sinh (\la + i \gamma S/2)}{\sinh (\la - i \gamma S/2)}
\]
The hyperbolic cosines in (\ref{bethesg}) cause no problem
because we can shift the $\la$'s
to convert them to  hyperbolic sines. The right hand side remains unchanged
since it depends only on differences of $\la$'s.
We conclude that the Sine-Gordon model can be viewed as inhomogeneous
spin (-1/2) chain.

Up to now we have consider only discretization in the space direction.
One can also consider discretization in the time direction.
There we have light-like Lax operators $\hat{L}$ which connect the
lattice points in the light-like direction (Fig. 10).

\setlength{\unitlength}{1mm}

\parbox[c]{130mm}{

\begin{picture}(130,40)

\put (65,10){\line(1,1){20}}
\put (80,20){$\hat{L}(\la-\omega)$}
\put (65,10){\line(-1,1){20}}
\put (35,20){$\hat{L}(\la-\omega)$}
\put (45,30){\line(1,0){40}}
\put (60,32){$L_n(\la,\omega)$}
\put (42,30){$n$}
\put (86,30){$(n+1)$}
\put (25,30){\vector(-1,0){10}}
\put (5,26){space direction}
\put (75,15){\vector(-1,-1){5}}
\put (60,10){\vector(-1,1){5}}
\put (76,9){light-like directions}
\end{picture}

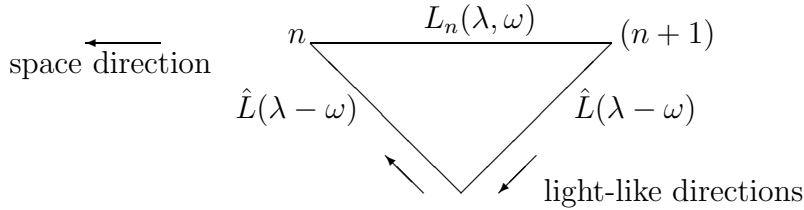
Figure 10. Discretization in the time direction.\\}

For further discussion we refer to the article \cite{14}.

This concludes these lectures. We hope, that we were able to show, that the
magnetic chains and the Bethe Ansatz contain a lot of potential for
classifying and solving the integrable models of quantum field theory.
The parameters, entering the description of magnetic chain, i.e. group $G$,
its representation and one (or two) anisotropy parameters
 can be identified with dynamical field variables,
coupling constants and mass parameter.
All known integrable models could be put into this scheme. The models of
conformal field theory then appear as a particular (massless) limit.


\begin{thebibliography}{99}

\bibitem{1} L. D. Faddeev, ``Integrable Models, Quantum Groups and
Conformal Field Theory'', Lectures in Technical University, Berlin 1992,
preprint no. SFB 288 N1.
\bibitem{2} L. D. Faddeev, ``The Bethe Ansatz'', Andrejewski Lectures at
Hamburg University, Berlin 1993, preprint no. SFB 288.
\bibitem{3} L. D. Faddeev, Lectures in Salamanka Summer School,
NATO ASI series, (1993), ed. Ibort, Plenum Press.
\bibitem{4} L. D. Faddeev, L. A. Takhtajan, ``Hamiltonian Methods in
Soliton Theory'', Springer, 1987.
\bibitem{5}  L. A. Takhtajan, L. D. Faddeev, Sov. J. Math. {\bf 24}, (1984) 241
\bibitem{6} D. Bernard, Int. J. Mod. Phys. {\bf B7}, (1993), 3517.
\bibitem{7} A. Izergin, V. E. Korepin, Lett. Math. Phys. {\bf 6}, (1982), 283.
\bibitem{8} V. Tarasov, L. A. Takhtajan, L. D. Faddeev, Theor. Math. Phys.
{\bf 57}, (1983) 163.
\bibitem{9} P. P. Kulish, N. Yu. Reshetikhin, Zaj. Nauch. Seminarov LOMI
{\bf 101}, (1981) 101, translation in J. Sov. Math. {\bf 23}, (1983) 2435.
\bibitem{10} V. Drinfield, ``Quantum Groups'' in Proceedings of ICM-86 at
Berkeley, vol. 1, 798, AMS 1987.
\bibitem{11} N. Yu. Reshetikhin, L. A. Takhtajan, L. D. Faddeev,
Leningrad Math. J. {\bf 1}, (1990) 193.
\bibitem{12} A. Yu. Alexeev, L. D. Faddeev, M. A. Semyonov-Tjan-Slansky,
CMP {\bf 149}, (1992) 335.
\bibitem{13} I. Frenkel, N. Yu. Reshetikhin, CMP {\bf 146}, (1992) 1.
\bibitem{14} L. D. Faddeev, A. Yu. Volkov, to be published in
Lett. Math. Phys.

\end{thebibliography}
\end{document}